\documentclass{JHEP}
\usepackage{graphicx}

\usepackage{amsmath}
\usepackage{amssymb}
                                           
\catcode`\@=11
\makeatletter \ifcase\@ptsize \font\teneufm=eufm10
\font\seveneufm=eufm7 \font\fiveeufm=eufm5
\font\teneusm=eusm10 \font\seveneusm=eusm7
\font\fiveeusm=eusm5 \or \font\teneufm=eufm10 scaled
\magstephalf \font\seveneufm=eufm7 \font\fiveeufm=eufm5
\font\teneusm=eusm10 scaled \magstephalf
\font\seveneusm=eusm7 \font\fiveeusm=eusm5 \or
\font\teneufm=eufm10 scaled \magstep1 \font\seveneufm=eufm7
\font\fiveeufm=eufm5 \font\teneusm=eusm10 scaled \magstep1
\font\seveneusm=eusm7 \font\fiveeusm=eusm5 \fi

\newfam\eufmfam \newfam\eusmfam \textfont\eufmfam=\teneufm
\scriptfont\eufmfam=\seveneufm
\scriptscriptfont\eufmfam=\fiveeufm
\textfont\eusmfam=\teneusm \scriptfont\eusmfam=\seveneusm
\scriptscriptfont\eusmfam=\fiveeusm

\def\frak{\ifmmode\let\next\frak@\else
 \def\next{\errmessage{Use \string\frak\space only in math
 mode}}\fi\next} \def\frak@#1{{\frak@@{#1}}}
 \def\frak@@#1{\fam\eufmfam#1} 
 \def\sh{\ifmmode\let\next\sh@\else
 \def\next{\errmessage{Use \string\sh\space only in math
 mode}}\fi\next} \def\sh@#1{{\sh@@{#1}}}
 \def\sh@@#1{\fam\eusmfam#1}

\ifcase\@ptsize \font\tenmsa=msam10 \font\sevenmsa=msam7
 \font\fivemsa=msam5 \font\tenmsb=msbm10
 \font\sevenmsb=msbm7 \font\fivemsb=msbm5 \or
 \font\tenmsa=msam10 scaled \magstephalf
 \font\sevenmsa=msam7 \font\fivemsa=msam5
 \font\tenmsb=msbm10 scaled \magstephalf
 \font\sevenmsb=msbm7 \font\fivemsb=msbm5 \or
 \font\tenmsa=msam10 scaled \magstep1 \font\sevenmsa=msam7
 \font\fivemsa=msam5 \font\tenmsb=msbm10 scaled \magstep1
 \font\sevenmsb=msbm7 \font\fivemsb=msbm5 \fi

\newfam\msafam \newfam\msbfam \textfont\msafam=\tenmsa
\scriptfont\msafam=\sevenmsa
\scriptscriptfont\msafam=\fivemsa \textfont\msbfam=\tenmsb
\scriptfont\msbfam=\sevenmsb
\scriptscriptfont\msbfam=\fivemsb

\def\Bbb{\ifmmode\let\next\Bbb@\else
 \def\next{\errmessage{Use \string\Bbb\space only in math
 mode}}\fi\next} \def\Bbb@#1{{\Bbb@@{#1}}}
 \def\Bbb@@#1{\fam\msbfam#1} \def\hexnumber@#1{\ifnum#1<10
 \number#1\else \ifnum#1=10 A\else\ifnum#1=11
 B\else\ifnum#1=12 C\else \ifnum#1=13 D\else\ifnum#1=14
 E\else\ifnum#1=15 F\fi\fi\fi\fi\fi\fi\fi}
 \def\msa@{\hexnumber@\msafam} \def\msb@{\hexnumber@\msbfam}
 \mathchardef\square="0\msa@03

\makeatother
\newcommand{\beq}{\begin{equation}}
\newcommand{\eeq}{\end{equation}}

\newcommand{\ba}{\begin{array}}
\newcommand{\ea}{\end{array}}
\newcommand{\bea}{\begin{eqnarray}}
\newcommand{\eea}{\end{eqnarray}}
\newcommand{\bean}{\begin{eqnarray*}}
\newcommand{\eean}{\end{eqnarray*}}

\newcommand{\be}{\begin{equation}}
\newcommand{\ee}{\end{equation}}

\newcommand{\ZZZ}{{\mathbb Z}} 

\newtheorem{theorem}{Theorem}[section]

\newtheorem{remark}[theorem]{Remark}

\newtheorem{proof}{Proof.}

\makeatother

\newcommand{\ZZ}{{\Bbb Z}}

 \def\be{\beta}

\def\be{\begin{equation}}
\def\ee{\end{equation}}
\def\l{\label}

\preprint{MIT-CTP-3372 \\ \\ {\tt
hep-th/}}

\title{Matrix Models, Argyres-Douglas singularities and double scaling limits}

\author{Gaetano Bertoldi
\footnote{
Research supported in part
by the CTP and the LNS of MIT and the U.S. Department of Energy 
under cooperative research agreement \# DE-FC02-94ER40818.
G. B. is also supported in part by the INFN ``Bruno Rossi''
Fellowship and by the Foundation BLANCEFLOR
Boncompagni-Ludovisi n\'ee Bildt.}
\\
Center for Theoretical Physics,
\\ Massachusetts Institute of Technology\\ Cambridge MA 02139\\
\email{bertoldi@mit.edu}
}

\abstract{
We construct an ${\cal N}=1$ theory with gauge group
$U(nN)$ and degree $n+1$ tree level superpotential
whose matrix model spectral curve develops
an $A_{n-1}$ Argyres-Douglas singularity.
We evaluate the coupling constants
of the low-energy $U(1)^n$ theory and show that the 
large $N$ expansion is singular at the Argyres-Douglas points.
Nevertheless, it is possible to define
appropriate double scaling limits 
which are conjectured to yield four dimensional 
non-critical string theories as proposed by
Ferrari. In the Argyres-Douglas 
limit the $n$-cut spectral curve  
degenerates into a solution with $\frac{n}{2}$ 
cuts for even $n$ and $\frac{n+1}{2}$ cuts for 
odd $n$. 
}


\begin{document}


\section{Introduction}

In \cite{DV1,DV2,DV3}, Dijkgraaf and Vafa 
conjectured that the
exact superpotential and gauge couplings of a class
of ${\cal N}=1$ super Yang-Mills
theories
can be calculated by doing perturbative
computations in an auxiliary matrix model. 
They considered theories 
with a polynomial superpotential ${\cal W}(\Phi)$ 
for the chiral adjoint field $\Phi$ and proposed that
${\cal W}$ is actually 
the potential in the related matrix model. 
Furthermore,
only planar diagrams in the matrix model contribute
to the effective superpotential.
This striking result
was later proved with perturbative field theory arguments 
in \cite{DGLVZ}
and by the analysis of the generalized Konishi anomaly in
\cite{CDSW}.
The solution of the matrix model in the planar limit is
captured by the so-called spectral curve, which
is given by
\be
y^2 = {\cal W}_n'(x)^2 + f_{n-1}(x)\,,
\l{spectre}\ee
where $n$ is the degree of ${\cal W}'(x)$ and
$f_{n-1}(x)$ is a polynomial of degree $n-1$.  
The above curve is a hyperelliptic Riemann surface 
of genus $n-1$, a double cover of
the $x$ plane with $2n$ branch points.
The values of the glueball superfields $S_k$ and the 
expression of the effective superpotential 
are related to integrals of the meromorphic
one-form $y\,dx$ over the curve (\ref{spectre}).

In \cite{Ferrari2}, Ferrari studied an ${\cal N}=1$ 
$U(N)$ gauge theory with cubic
superpotential and using the results of
\cite{Ferrari1} discovered that, in the phase where the
gauge group is unbroken, there are critical values 
of the superpotential couplings where
the effective superpotential is non-analytic and 
the standard large $N$ expansion is singular,
namely its coefficients are divergent. 
In fact, the tension of supersymmetric domain walls 
scales as a fractional power of $N$ 
at the critical points. This breakdown of the
$1/N$ expansion can actually be compensated by taking the
limit $N \to \infty$ and approaching the critical 
points in a correlated way. Furthermore,
these {\it double scaling limits} are conjectured 
to define a four dimensional non-critical string theory.
This relies on a proposal made by Ferrari
on how to generalize the old matrix model
approach to non-critical strings \cite{2dMMNCST}
to the four dimensional case 
\cite{FerrariNCST,FerrariNPB612,FerrariNPB617,FerrariNCSTlast,FerrariLH}.

It was also shown in \cite{Ferrari2} that, 
from the matrix model point of view,
the singularity corresponds to a transition
from a two-cut solution to a one-cut solution.
A cycle of the genus $1$ spectral curve that describes 
the two-cut solution shrinks to zero size.
In field theory language, this is a contact point
between two different patterns of gauge symmetry breaking. 
Ferrari also raised the question of the structure of higher order
critical points \`a la Argyres-Douglas \cite{AD}. 
A matrix model spectral curve undergoes such 
a degeneration when two or more cycles with 
{\it non-vanishing intersection} 
shrink to zero size simultaneously.
These singularities were first investigated
in the context of ${\cal N}=2$ super Yang-Mills
theories, whose low-energy physics
is encoded by Seiberg-Witten hyperelliptic curves
\cite{SW,Ketal,AF}. 
Their importance lies in the fact that,
since the vanishing cycles have non-trivial
intersection, the low-energy
theory contains both electric and magnetic charges \cite{AD}.
Furthemore, they are non-trivial interacting ${\cal N}=2$
conformal field theories 
\cite{AD,APSW,Eguchi1,Eguchi2}
and they provided the first quantitative check 
of the scenario advocated by Ferrari \cite{FerrariNPB617}.

In this paper, higher critical points \`a la Argyres-Douglas
in ${\cal N}=1$ theories are constructed and studied.
In particular, a $U(nN)$ gauge theory breaking to $U(N)^n$
in the presence of the one-parameter superpotential 
\be
{\cal W}(\Phi) = g_n \left(
\frac{1}{n+1} \,\Phi^{n+1} - u\,\Phi
\right)\,, \quad n \geq 3\,,
\l{maremma}\ee
is analysed in detail. There are two values
of the parameter $u$ where the spectral curve
develops an $A_{n-1}$ Argyres-Douglas singularity.

The plan of the paper is as follows.
In section \ref{AD},  
Argyres-Douglas singularities are briefly 
reviewed.
In section \ref{strong}, 
the {\it strong coupling approach} 
to the study of softly broken ${\cal N}=1$ 
theories \cite{CIV} is also reviewed. 
This is the essential instrument to
engineer models whose spectral curve
develops an Argyres-Douglas singularity {\it on-shell}.
The choice
of the particular model (\ref{maremma})
and the reason why it is expected to develop
an Argyres-Douglas singularity are thus explained.

\noindent
In section \ref{glueballs}, the values of the glueball
superfields $S_k$, $k=1, \ldots n$,
are calculated exactly by first showing that they
are solutions of a linear second-order differential
equation in $u$ and then evaluating
them in the semiclassical limit. 
They are non-analytic at the Argyres-Douglas points.
The differential equation is the Picard-Fuchs equation 
for the periods of the meromorphic one-form $ydx$ on the 
spectral curve. 

\noindent
In section \ref{multiN}, the {\it multiplication map} by $N$
\cite{CIV} is used to map 
the original $U(n)$ theory breaking to $U(1)^n$ 
to a $U(nN)$ theory breaking to $U(N)^n$ with the same
superpotential (\ref{maremma}), 
thereby enabling us to take the large $N$ limit. 
The single vacuum of the $U(n)$ theory 
corresponding to the above symmetry breaking
is mapped to $N$ vacua of the $U(nN)$ theory.

\noindent
In section \ref{Weff}, it is shown that the effective 
superpotential for all the vacua is vanishing.
In section \ref{MatrixAnalysis}\,, 
it is proved that the ansatz for the
spectral curve satisfies the matrix model equations
of motion consistent with the above symmetry
breaking pattern, $U(nN) \rightarrow U(N)^n$.
In particular, the
results found via the {\it multiplication map} 
are reproduced.  
This paves the way for the evaluation of
the coupling constants of the low-energy
$U(1)^n$ theory in section \ref{couplings}. 
They are non-analytic at the critical points. 
Furthermore, in the Argyres-Douglas 
limit the $n$-cut spectral curve  
degenerates into a solution with $\frac{n}{2}$ 
cuts for even $n$ and $\frac{n+1}{2}$ cuts for 
odd $n$. 
Finally, in section \ref{largeN}, it is shown
that the large $N$ expansion 
is singular at the Argyres-Douglas points,
with the coupling constants scaling in general as 
a fractional power of $N$. However, as in \cite{Ferrari2},
there exists a well--defined double scaling limit
\be
x \to 1 \,, \quad N \to \infty\,, 
\quad  N(1-x) = cnst = \frac{1}{\kappa}\,,
\quad x = \frac{4\Lambda_0^{2n}}{u^2}\,,  
\label{doubleintro}\ee
which is conjectured to yield 
a non-perturbative definition 
of a four dimensional
non-critical string theory.

\section{Argyres-Douglas singularities}\label{AD}

Argyres-Douglas singularities were 
originally investigated in the 
context of ${\cal N}=2$ super Yang-Mills
theories, whose low-energy physics
is encoded by Seiberg-Witten hyperelliptic curves
\cite{SW,Ketal,AF}.
The fact that the vanishing cycles have non-trivial
intersection implies that the low-energy ${\cal N}=2$ theory 
has massless solitons with {\it mutually non-local} charges \cite{AD}.
Namely, these solitons are both electrically and magnetically
charged under the same $U(1)$ factor.
The theories at these points are actually 
superconformal \cite{APSW}.
As an illustration,
let us consider the ${\cal N}=2$ Seiberg-Witten curve for $SU(3)$
\cite{SW,Ketal,AF}
$$
y^2 = (x^3 - u x - v)^2 - 4 \Lambda^6\,.
$$  
The above genus $2$ hyperelliptic curve 
is singular whenever the polynomial on the r.h.s.
has at least a double root, which is 
equivalent to the vanishing of its discriminant 
$$
\Delta =2^{12} \Lambda^{18} \left( 4 u^3 - 27(v+2\Lambda^3)^2 \right)
\left( 4 u^3 - 27(v-2\Lambda^3)^2 \right)\,.
$$
For instance for $v=0$, $u = 3\, e^{2\pi i k/3} \Lambda^2$, $k=0,1,2$, 
the curve reduces to
$$
y^2 = ( x - e^{\pi i k/3} \Lambda )^2 ( x + e^{\pi i k/3} \Lambda )^2
( x^2 - 4 e^{2\pi i k/3} \Lambda^2 )\,,  
$$
which has two double roots. This is the limit where two
{\it mutually local} dyons become massless. 
Argyres-Douglas points in moduli space, however, correspond  
to higher order singularities. An example is given by $u=0$ 
and $v = \pm 2 \Lambda^3$, where the curve becomes \cite{AD}
\be
y^2 = x^3 ( x^3 - 2 v )\,,
\label{AD1}\ee
and correspondingly two cycles with non-vanishing intersection
shrink to zero size. 

In the following, a simple generalization of the
above singularity will be considered.
In particular, the on shell spectral curve 
of the ${\cal N}=1$ system studied in the paper 
is going to be
$$
y^2 = (x^n - u)^2 - 4 \Lambda^{2n}\,, \quad n \geq 3\,.
$$
It is easy to recognize that for $u = \pm 2\Lambda^n$,
$n$ out of the $2n$ branch points coalesce leading to an
$A_{n-1}$ Argyres-Douglas singularity
\be
y^2 \sim x^n\,.
\l{A(n-1)}\ee
Before introducing the specific model which is 
object of study, 
it is necessary to review the strong coupling
approach to the study of ${\cal N}=1$ 
gauge theories with polynomial 
superpotentials ${\cal W}(\Phi)$ \cite{CIV,BoerOz}.

\section{The strong coupling approach}\label{strong}

The dynamics of ${\cal N}=1$ $U(N)$ gauge theories
with polynomial superpotentials 
can be studied by treating 
${\cal W}(\Phi)$ as a perturbation of the 
underlying strongly coupled gauge theory
with ${\cal W}=0$. The latter system has 
${\cal N}=2$ supersymmetry and a Coulomb
moduli space of vacua described 
by a Seiberg-Witten curve 
\cite{SW,Ketal,AF}
$$
y^2 = P_N^2(x) - 4 \Lambda^{2N}\,,
$$
where the coefficients of $N$-th order 
polynomial $P_N(x)$ 
depend on the $N$ moduli 
$\langle tr \Phi^r \rangle,\, r = 1,\dots,N$.
In this strong coupling approach, which was
developed in \cite{CIV} 
using the methods of \cite{BoerOz}, 
${\cal W}$ is regarded as an effective
superpotential on the moduli space.
The generic low energy group on the
Coulomb moduli space is $U(1)^N$.
Vacua in which the low energy group of
the ${\cal N}=1$ theory is  
$U(1)^n$, for $n < N$, can be found by extremizing the 
superpotential on submanifolds of the 
Coulomb branch where $N-n$ monopoles of the
${\cal N}=2$ theory are massless.
The superpotential lifts all of the moduli
space except for a finite set of vacua.
At points where $N-n$ mutually local monopoles
become massless, the Seiberg-Witten 
curve has the following factorization
\be
y^2 = P_N(x)^2 - 4 \Lambda^{2N} = F_{2n}(x) H^2_{N-n}(x)\,,
\l{sca1}\ee
where the polynomials on the r.h.s.\,\,have simple roots.
This factorization is satisfied on an $n$-dimensional
submanifold of the Coulomb moduli space on which
the superpotential should be extremized in order to
find the ${\cal N}=1$ vacua. In \cite{CIV},
Cachazo, Intriligator and Vafa showed that this 
yields an on shell relation between the tree level 
superpotential and the polynomial $F_{2n}(x)$.
In particular, when the degree of ${\cal W}'(x)$
is equal to $n$, the highest $n+1$ coefficients of
$F_{2n}(x)$ are given in terms of ${\cal W}'(x)$ 
as follows 
\be
F_{2n}(x) = \frac{1}{g_n^2} {\cal W}'(x)^2 
+ {\cal O}(x^{n-1})\,.
\l{sca2}\ee
Given ${\cal W}$, the above relation determines $F_{2n}(x)$ 
in terms of $n$ unknown coefficients that are fixed 
by requiring the existence of a polynomial 
$H_{N-n}(x)$ such that the factorization (\ref{sca1})
holds. This determines $P_N(x)$ or equivalently
the ${\cal N}=2$ vacuum.

Conversely, given a polynomial $P_N(x)$ with
the above factorization, one may look for 
a superpotential consistent with this vacuum.
This inverse technique was used in \cite{CV}
to rederive the ${\cal N}=2$ solution
using the geometric engineering approach 
of \cite{CIV}. It was also used in \cite{CSW}
to study the various phases of such ${\cal N}=1$
gauge theories and the structure of their parameter
space. The same authors also provided
a generalization of the above approach
to include the cases $deg\,{\cal W}(\Phi) > n+1$ and 
$deg\,{\cal W}(\Phi) > N$.  

Therefore, in order to construct examples
of matrix model spectral curves which
develop Argyres-Douglas singularities,
one can start from a 
$p$-parameter family of ${\cal N}=2$
hyperelliptic curves that displays 
such a degeneration in some appropriate limit. 
Then, using the inverse technique,
it is possible to determine a superpotential
consistent with these curves.
The procedure yields the corresponding
on shell family of ${\cal N}=1$ spectral curves 
and once this is
given one can study the gauge theory along
the lines of \cite{CIV,DV1,DV2,DV3}. 
In the following, a one-parameter family
of genus $n-1$ hyperelliptic curves that
can develop Argyres-Douglas singularities
is introduced.
Then, a consistent order $(n+1)$ 
superpotential ${\cal W}(\Phi)$
is determined.  
Finally, the effective superpotential,
glueball superfields $S_k$ and coupling constants
of the low-energy abelian gauge theory
will be evaluated.


\subsection{The model}

Consider a $U(n)$ Seiberg-Witten curve
of the following form
\be
y^2 = P_n(x)^2 - 4 \Lambda^{2n} 
= \left(\, x^n - u \,\right)^2 - 4\Lambda^{2n} \,.
\l{SW1}\ee
The above curve has genus $n-1$ and is singular whenever the
discriminant $\Delta$ of the polynomial on the right hand side
of (\ref{SW1}) vanishes
\be
\Delta = (2n)^{2n} ( - 4 \Lambda^{2n} )^n ( u^2 - 4 \Lambda^{2n} )^{n-1} \,.
\label{Delta}\ee
In particular, in the limit $u \to \pm 2 \Lambda^n$,
$n$ branch points collide and the curve reduces to
$$
y^2 = x^n \left(\, x^n \mp 4 \Lambda^{n} \,\right)\,.
$$
For $n \geq 3$, these are Argyres-Douglas singularities. 
Note that the curve $(\ref{SW1})$ depends on one parameter only, $u$,
and that it has a $\ZZZ_n$ symmetry generated by
\be
\left( x , y \right) \to \left( e^{2\pi i/n} x , y \right)\,.
\label{Zn}\ee
In this case the factorization of the 
Seiberg-Witten curve is trivial, namely for 
a generic value of $u$ there are no double 
roots   
\be
y^2 = P_n^2(x) - 4 \Lambda^{2n} = F_{2n}(x)\,.
\l{model1}\ee
Then, the low energy group of the above theory is $U(1)^n$
and the degree of a polynomial superpotential 
${\cal W}(\Phi)$ consistent with the above Seiberg-Witten curve 
has to be at least $n+1$.
When the degree of ${\cal W}(\Phi)$ is equal to $n+1$, 
the matrix model spectral curve actually coincides with 
the Seiberg-Witten curve \cite{CV}
\be
y^2 = P_n^2(x) - 4 \Lambda^{2n} = F_{2n}(x) = 
\frac{1}{g_n^2} \left( 
{\cal W}_n'(x)^2 + f_{n-1}(x)
\right)
\,,
\label{model2}\ee
where
\be
{\cal W}'_n(x) = g_n P_n(x)\,, \quad f_{n-1}(x) = - 4\,g^2_n \Lambda^{2n}\,.  
\label{model3}\ee 
Then, modulo the addition of a constant, the tree level superpotential is
\be
{\cal W}(\Phi) = \frac{g_n}{n+1}\,\Phi^{n+1} - g_n u\,\Phi\,.
\label{super1}\ee
When $u=0$, the field $\Phi$ becomes critical and there is
a singularity in the classical space of parameters. 
Note also that, since $f_{n-1}(x)$ is constant, 
the total glueball superfield 
$S = \sum_{k=1}^n S_k$ vanishes identically \cite{CV,CDSW}.
Finally, by the {\it multiplication map} introduced in \cite{CIV},
the above $U(n)$ theory can be mapped to a
$U(nN)$ theory with the same superpotential.
In the remainder of the paper we are going to set $g_n = 1$.

\vspace{1.0cm}
\begin{figure}[ht]
\begin{center}
\input{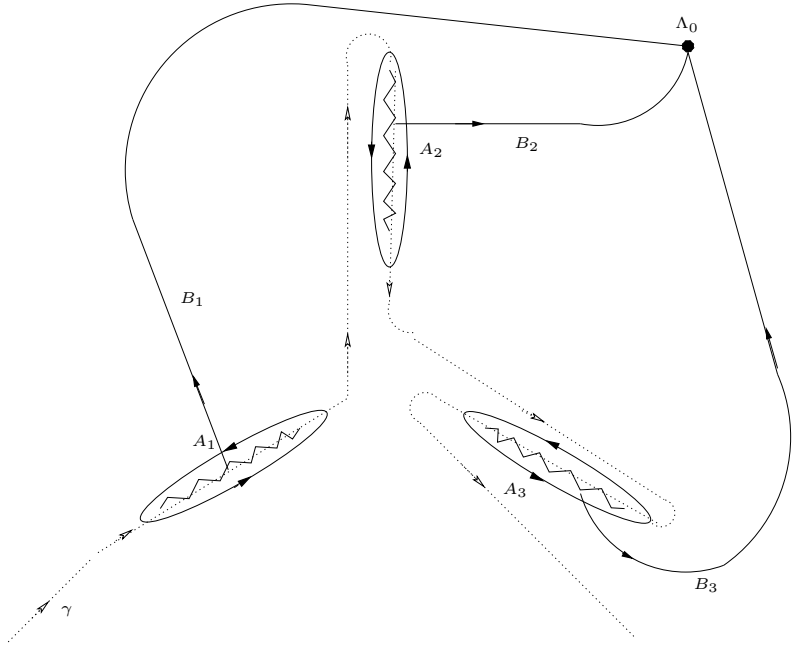}
\caption{ \label{fig:ABLambda_0} The $A$ and $B$ cycles 
for $n=3$, $( Im\,u = 0 )$.
The path $\gamma$ enters in the rigorous definition
of the matrix model conjecture recently studied by
Lazaroiu \cite{Laza} involving holomorphic
matrix models. By definition, $\gamma$ 
threads the branch cuts and fixes the basis
of $A$ and $B$ cycles on the spectral curve.
}
\end{center}
\end{figure}
\vspace{0.5cm}

\section{The glueball superfields}\label{glueballs}


\noindent
The glueball superfields $S_k$ are given by the period integrals
of the meromorphic one-form $y dx$ along the closed
loops $A_k$ surrounding the $k$-th branch cut \cite{CIV,DV1,DV2,DV3}
\be
S_k = \frac{1}{2 \pi i} \oint_{A_k} y \,dx = \frac{1}{2 \pi i} 
\oint_{A_k} \sqrt{\left( x^n - u \right)^2 - 4\Lambda^{2n}} \,dx\,.
\l{Si}\ee
We are going to find the exact expression of the above periods
by first deriving a second-order linear differential equation in $u$
satisfied by them, a so-called Picard-Fuchs equation. 
The evaluation of the semiclassical limit
of $S_k$ will then allow to fix the linear combination
of the two independent solutions.
A similar analysis was carried out in the context of ${\cal N}=2$
Seiberg-Witten theories in \cite{KLT}.
First of all
$$
\partial_u S_k = \frac{1}{2 \pi i} \oint_{A_k} \partial_u (\,y \,dx\,) 
= - \frac{1}{2 \pi i} \oint_{A_k} \frac{( x^n - u)}{y} dx\,.
$$
Taking another derivative we find
$$
\partial^2_u S_k = \frac{1}{2 \pi i} \oint_{A_k} \partial_u^2 ( \,y\,dx\,) = 
- \frac{1}{2 \pi i} \oint_{A_k} \frac{4 \Lambda^{2n}}{y^3} dx\,.
$$
By the following identity
$$
\partial_u^2 ( \,y\,dx\,) = \left( \frac{n-1}{n} \right)
\frac{4 \Lambda^{2n}}{4 \Lambda^{2n}-u^2} \frac{dx}{y} 
+ \frac{(x^n-u)u}{n (4 \Lambda^{2n} - u^2)}
\frac{dx}{y} + \partial_x \left( 
\frac{x(u^2 - u x^n + 4 \Lambda^{2n})}{n(4\Lambda^{2n}-u^2)y}
\right) dx\,,
$$
$$
= - \left( \frac{n-1}{n^2} \right)
\frac{1}{4 \Lambda^{2n}-u^2} 
\Lambda \partial_\Lambda ( y dx ) 
- \frac{1}{n} \frac{u}{(4 \Lambda^{2n} - u^2)}
\partial_u ( y dx ) 
$$
$$
+ \partial_x \left( 
\frac{x(u^2 - u x^n + 4 \Lambda^{2n})}{n(4\Lambda^{2n}-u^2)y}
\right) dx\,,
$$
we conclude that
\be
\partial^2_u S_k = \left( \frac{n-1}{n^2} \right)
\frac{1}{u^2 - 4 \Lambda^{2n}} 
\Lambda \partial_\Lambda S_k 
+ \frac{1}{n} \frac{u}{(u^2 - 4 \Lambda^{2n})}
\partial_u S_k\,, 
\label{fast1}\ee
since the integral of an exact differential along a closed cycle is zero.

By (\ref{Si}), we can also see that $S_k$ is a homogeneous function
of degree $n+1$, which yields
\be
\left( \Lambda \partial_\Lambda + n u \partial_u \right) S_k
= (n+1) S_k\,.
\label{homog}\ee
Finally, by (\ref{fast1}) and (\ref{homog}), we find 
\be
\left[ \partial_u^2  
+ \left( \frac{n-2}{n} \right) \frac{u}{u^2-4\Lambda^{2n}}
\partial_u - \left( \frac{n^2-1}{n^2} \right) \frac{1}{u^2-4\Lambda^{2n}}
\right] S_k = 0\,. 
\label{fastfinal}\ee
Note that every period of $y\,dx$ 
is a solution of the above Picard-Fuchs equation
as long as the cycle is closed.  
In particular, the difference of any two derivatives
of the prepotential ${\cal F}$ with respect to the 
glueball superfields is also a solution of (\ref{fastfinal}).

\subsection{Solution of the Picard-Fuchs equation}

By the following change of variables 
$$
z = \frac{u^2}{4\Lambda^{2n}}
$$
Eq.(\ref{fastfinal}) becomes a hypergeometric equation
\be
\left[ z(1-z) \partial^2_z 
+ \left(\frac{1}{2} - \frac{n-1}{n} z \right) \partial_z 
+ \frac{n^2-1}{4 n^2} 
\right] S_k = 0\,,
\label{PFhyper}\ee
which is solved by \cite{Erdelyi}
\be
C_{1,k}\, F\left( \frac{1}{2} -\frac{1}{2n}, -\frac{1}{2} -\frac{1}{2n}, 
\frac{1}{2}, \frac{u^2}{4\Lambda^{2n}} 
\right) 
+ C_{2,k}\, u\, F\left( 1 - \frac{1}{2n}, -\frac{1}{2n}, \frac{3}{2},
\frac{u^2}{4\Lambda^{2n}} \right)\,.
\label{Szero}\ee
\noindent
In order to fix the coefficients $C_{1,k}$ and $C_{2,k}$,
one can evaluate $S_k$ in the semiclassical limit $\Lambda \to 0$,
where the glueball superfield vanishes.
Since $\lim_{\Lambda \to 0} z = \infty$, one needs to
perform the analytic continuation of the above
hypergeometric functions, which are 
defined as power series in the disc $|z| \leq 1$. 
Alternatively, one can rewrite the Picard-Fuchs equation 
in terms of a new variable that vanishes as $\Lambda \to 0$.

\noindent
In the limit $u \to \infty$, which is dual to $\Lambda \to 0$,
the solutions of Eq.(\ref{fastfinal}) are asymptotic to
$$
u^{\alpha_{\pm}} f_{\pm}(u)\,,
$$
where
$$
\alpha_{\pm} = \frac{1}{n} \pm 1\,, \quad
\lim_{u \to \infty} f_{\pm}(u) \ne 0\,.
$$
Setting $S_k = u^{\alpha_{-}} f_k(u)$ and changing variables to
$z = \frac{4\Lambda^{2n}}{u^2}$, Eq.(\ref{fastfinal})
is equivalent to
\be
\left[ z(1-z) \partial^2_z 
+ \left( 2 - \left( \frac{5}{2}-\frac{1}{n} \right) z \right) \partial_z 
- \frac{2n^2 - 3n + 1}{4 n^2}  
\right] f_k = 0\,,
\label{EQutile}\ee 
which is again a hypergeometric equation.
The condition that $S_k$ vanishes in the limit 
$z = \frac{4\Lambda^{2n}}{u^2} \to 0$ implies that
\be
S_k = C_{3,k}\,u^{-1 + \frac{1}{n}} \,
F\left(
\frac{1}{2}-\frac{1}{2n},1 - \frac{1}{2n},2,\frac{4\Lambda^{2n}}{u^2} 
\right).
\l{Sinfinity}\ee
Note that the above expression is a power series
in $\Lambda^{2n}$, namely $S_k$ is
given by an instanton sum.
The value of $C_{3,k}$ can be found by evaluating
the semiclassical limit of $S_k$ more carefully
\be
S_k = \frac{1}{2 \pi i} \oint_{A_k} \sqrt{ P_n^2 - 4\Lambda^{2n} }\,dx 
= - \frac{ 2\Lambda^{2n} }{2 \pi i} \oint_{\gamma_k} \frac{1}{P_n(x)} \,dx\,
+ {\cal O}(\Lambda^{4n})\,, 
\label{res}\ee
where $\gamma_k$ is a counterclockwise loop around
the $k$-th root of $P_n(x)=x^n - u$,
$$
x_k = e^{2 \pi i k/n}  u^{1/n}\,. 
$$ 
Then
$$ 
-\frac{2 \Lambda^{2n}}{2 \pi i} \oint_{\gamma_k} \frac{1}{P_n(x)} \,dx\,
= - 2 \Lambda^{2n} \frac{1}{P'(x_k)} 
=  - \frac{2 \Lambda^{2n}}{n} e^{2 \pi i k/n} u^{-\frac{n-1}{n}}\,. 
$$
Finally, by (\ref{Sinfinity}), we find
\be
C_{3,k} =  - \frac{2 \Lambda^{2n}}{n}
e^{2\pi i k/n} \,,
\label{C3new}\ee
and
\be
S_k =  - \frac{2 \Lambda^{2n}}{n} e^{2\pi i k/n} u^{-1+1/n}
\,F\left(\frac{1}{2}-\frac{1}{2n},1 -\frac{1}{2n},2,\frac{4\Lambda^{2n}}{u^2} \right)
\,.
\label{Skcrismi}\ee
Note that this is consistent with the fact
that
$$
\sum_{k=1}^n S_k = 0\,,
$$
since $\sum_{k=1}^n e^{2 \pi i k/n} = 0$.
Actually, $S_{k+1} =
e^{2 \pi i/n} S_k$ as a direct consequence of the
symmetry (\ref{Zn}). 
In fact
$$
S_k = \frac{1}{2 \pi i} \oint_{A_k} y dx 
=  \frac{2}{2 \pi i} \int_{x_{k,-}}^{x_{k,+}} y dx\,,
$$
where
$$
x_{k,\pm} = e^{2\pi i k/n} \left( u \pm 2 \Lambda^{n} \right)^{1/n}\,,
$$
are the branch points of the spectral curve (\ref{SW1})
and by the change of variable
$x = e^{2\pi i/n} \,\tilde x$, we find
\be
S_k = \frac{2}{2 \pi i} \,e^{2\pi i/n} \int_{x_{k-1,-}}^{x_{k-1,+}} 
y d\tilde x\, = e^{2 \pi i/n} S_{k-1}\,.
\label{SkSk-1}\ee
Similarly, we can show that 
$\frac{\partial {\cal F}}{\partial S_k}
= e^{2\pi i (k-1)/n} \frac{\partial {\cal F}}{\partial S_1}$.

\noindent
Performing the analytic continuation of (\ref{Szero}),
which defines an absolutely convergent series for
$|u| \leq 2\Lambda^n$, to the region $|u| > 2\Lambda^n$, 
where (\ref{Sinfinity}) is valid, we are going to find relations
among the various $C_k$'s. The details are given in the Appendix.
The results are
\be
C_{2,k} = -\frac{e^{i\pi/2}}{2 \Lambda^n} 
\frac{\Gamma(1-1/2n)\Gamma(3/2+1/2n)}{\Gamma(1/2 - 1/2n)\Gamma(1+1/2n)}
\,C_{1,k}\,.
\label{C2C1}\ee
and
\be
C_{3,k} = e^{i\pi/2 + i\pi/2n} \frac{2 \pi}{\sin(\pi/n)}\,
\left( 
\frac{\sqrt{\pi}\,2^{\frac{n-1}{n}} }{2n \Gamma(1 + 1/2n)\Gamma(-1/2-1/2n)}
\right) \Lambda^{n-1} C_{1,k}\,.
\label{C3C1}\ee

\subsection{Non-analytic behaviour close to the Argyres-Douglas points}

By analytic continuation of (\ref{Sinfinity})(\ref{Skcrismi}), we find 
\cite{Erdelyi}
$$
S_k = C_{3,k}\,u^{-1 + \frac{1}{n}} \,
\left(
A_1 F\left(\frac{1}{2}-\frac{1}{2n},1-\frac{1}{2n},\frac{1}{2}
-\frac{1}{n},1-\frac{4\Lambda^{2n}}{u^2}\right)
\right.
$$
\be
\left.
+ A_2 ( 1 - 4\Lambda^{2n}/u^2 )^{\frac{n+2}{2n}}
F\left( \frac{3}{2}+\frac{1}{2n},1+\frac{1}{2n},\frac{3}{2}
+\frac{1}{n},1-\frac{4\Lambda^{2n}}{u^2}\right)
\right)\,,
\label{Su=1}\ee
which implies that close to the singularity at $u^2 = 4\Lambda^{2n}$
\be
S_k \approx A_1 
+ A_2 
\left( 
\frac{u^2 - 4\Lambda^{2n}}{4\Lambda^{2n}} 
\right)^{\frac{n+2}{2n}} + {\cal O}(u^2 - 4\Lambda^{2n})\,,
\label{ADasymp}\ee
where
\be
A_1 = \frac{\Gamma(1/2+1/n)}{\Gamma(3/2+1/2n)\Gamma(1+1/2n)}\,,\quad
A_2 = \frac{\Gamma(-1/2-1/n)}{\Gamma(1/2-1/2n)\Gamma(1-1/2n)}\,.
\label{A1A2old}\ee
Thus, we see that $S_k$ is non-vanishing in the limit $u^2 \to 4\Lambda^{2n}$ 
and that it is non-analytic due to the fractional exponent $\frac{n+2}{2n}$, 
$( n \geq 3 )$.


\section{The multiplication map}\label{multiN}

By the so-called {\it multiplication map} by $N$ 
introduced in \cite{CIV},
the above $U(n)$ theory with superpotential (\ref{super1})
can be mapped to a $U(nN)$ theory 
with the same tree level superpotential. In particular, the vacuum 
considered up to now,
which is a Coulomb vacuum with unbroken $U(1)^n$ gauge group,
is associated to $N$ different vacua with unbroken $U(N)^n$.
In fact, given a set of polynomials $P_n(x)$, $F_{2m}(x)$ and
$H_{n-m}(x)$, all with the highest coefficient equal to $1$,
that satisfy the following relations
\be
P_{n}^2(x) - 4 \Lambda^{2n} = F_{2m}(x) H^2_{n-m}(x)\,,
\quad
F_{2m}(x) = {\cal W}'(x){}^2 + f_{m-1}(x)\,,  
\l{multi1}\ee
it is possible to show that 
${\cal Q}_{nN}(x) \equiv 2 \Lambda_0^{nN} \eta^{N} {\cal T}_N \left( 
\frac{P_n(x)}{2 \eta \Lambda_0^n} 
\right)$, where ${\cal T}_N$ is the $N$-th Chebishev polynomial of the first kind
and $\eta$ is a $2N$-th root of unity,
is a polynomial of degree $nN$ with highest coefficient 
equal to $1$ and that it satisfies
$$
{\cal Q}^2_{nN}(x) - 4 \Lambda_0^{2nN} = 
4 \Lambda_0^{2nN} \left(
{\cal T}^2_N \left( 
\frac{ P_n(x) }{2 \eta \Lambda_0^n} 
\right) - 1 
\right)  
$$
$$
= 4 \Lambda_0^{2nN} \left[
\left(  \frac{ P_n(x) }{2 \eta \Lambda_0^n} \right)^2 
- 1 \right] {\cal U}^2_{N-1} \left( 
\frac{ P_n(x) }{2 \eta \Lambda_0^n} 
\right)
$$
$$
= \left( P_n^2(x) - 4 \eta^2 \Lambda_0^{2n} \right) 
\left( \eta^{N-1} \Lambda_0^{n(N-1)}\,{\cal U}_{N-1} \left( 
\frac{ P_n(x) }{2 \eta \Lambda_0^n} 
\right)
\right)^2\,,
$$
where ${\cal U}_{N-1}$ is the $(N-1)$-th Chebishev polynomial
of the second kind and we used the following relation
\be
{\cal T}^2_N(x) - 1 = ( x^2 - 1 ) {\cal U}^2_{N-1}(x)\,.   
\l{cheb1}\ee
Therefore, if one sets
\be
\Lambda^{2n} = \eta^2 \Lambda_0^{2n}
= e^{2 \pi i p/N} \Lambda_0^{2n}
\,, \quad p=0,\ldots,N-1\,, 
\l{scalesN}\ee
one finds that ${\cal Q}_{nN}(x)$ 
satisfies the following identity
$$
{\cal Q}^2_{nN}(x) - 4 \Lambda_0^{2nN} 
= F_{2m}(x) H^2_{n-m}(x)
\left( \eta^{N-1} \Lambda_0^{n(N-1)}\,{\cal U}_{N-1} \left( 
\frac{ P_n(x) }{2 \eta \Lambda_0^n} 
\right)
\right)^2
$$
\be
= \tilde F_{2m}(x) \tilde H^2_{nN-m}(x)\,,
\l{mapN1}\ee
where
\be
\tilde F_{2m}(x) = F_{2m}(x)\,,
\quad \tilde H_{nN-m}(x) = 
H_{n-m}(x)\, \eta^{N-1} \Lambda_0^{n(N-1)}\,
{\cal U}_{N-1} \left( 
\frac{ P_n(x) }{2 \eta \Lambda_0^n} 
\right)
\l{mapN2}\ee
As was explained in section \ref{strong}, 
if one wants to find vacua of a $U(nN)$ theory 
with low energy group $U(1)^m$, one has to solve the 
following factorization problem
$$
P^2_{nN}(x) - 4 \Lambda_0^{2nN} = \tilde F_{2m}(x) 
\tilde H^2_{nN-m}(x)\,.
$$
Then, Eq.(\ref{mapN1}) implies that $P_{nN}(x)={\cal Q}_{nN}(x)$
is a solution. Furthermore, since $\tilde F_{2m}(x) = F_{2m}(x)$,
the vacua of the $U(nN)$ theory have the same superpotential
as the vacua of the $U(n)$ theory.
It was also shown in \cite{CIV}, that a classical
limit with unbroken $\prod_{j=1}^k U(n_j)$ is mapped to a classical limit 
with unbroken $\prod_{j=1}^k U( N n_j)$.
Finally, Eq.(\ref{scalesN}) implies that 
for each vacuum of the $U(n)$ theory one has
$N$ vacua of the $U(nN)$ theory.

By Eqs.(\ref{model2}),(\ref{scalesN}) and (\ref{mapN2}), 
the spectral curve relative to one of
the $N$ vacua of the $U(nN)$ theory breaking to
$U(N)^n$ is given by
\be
y^2 = \left( x^n - u \right)^2 - 4\,e^{2 \pi i p/N} \Lambda_0^{2n}
\,, \quad p=0,\ldots,N-1\,.
\l{spectralnN}\ee
This implies that the glueball superfields 
$S_k,\,k=1,\ldots,n$ of each $U(N)$ factor are given by
\be
S_k =  - \frac{2\,e^{2 \pi i p/N} \Lambda_0^{2n}}{n} e^{2\pi i k/n} u^{-1+1/n}
\,F\left(\frac{1}{2}-\frac{1}{2n},1 -\frac{1}{2n},2,
\frac{4\,e^{2 \pi i p/N} \Lambda_0^{2n}}{u^2} \right)\,. 
\label{SkcrismiN}\ee

\section{The effective superpotential}\label{Weff}

Since the glueball superfield $S = \sum_{k=1}^n S_k$
vanishes exactly, the effective superpotential
is given by the classical expression.
In fact
$$
\Lambda^{2nN} \frac{\partial}{\partial \Lambda^{2nN}} {\cal W}_{eff} 
= \langle S \rangle = 0\,,
$$ 
which implies that there are no quantum corrections to ${\cal W}_{eff}$.
The superpotential (\ref{maremma}) has $n$ classical ground states 
where
$$
\langle \Phi \rangle = e^{2 \pi i k/n} u^{\frac{1}{n}}\,,
\quad k=1,\ldots,n\,.
$$
Then, for the $U(nN)$ theory breaking to
$U(N)^n$ we find that
$$
{\cal W}_{eff} = {\cal W}_{cl} =
g_n \left(
\frac{1}{n+1} \, \langle \Phi^{n+1} \rangle 
- u\, \langle \Phi \rangle 
\right)
$$
$$
= g_n \left(
\frac{1}{n+1} \, 
\sum_{k=1}^n 
e^{2 \pi i k/n} u^{(n+1)/n}  
- u\, \sum_{k=1}^n e^{2 \pi i k/n} u^{1/n} 
\right) = 0\,.
$$


\section{The Matrix Model analysis}\l{MatrixAnalysis}

In this section, we are going to show
that the ansatz (\ref{model2}),(\ref{model3}),(\ref{super1}) 
for the spectral curve indeed satisfies the matrix model 
equations of motion consistent with the 
symmetry breaking pattern 
$U(nN) \to U(N)^n$. In particular,
we are going to find $N$ different vacua,
characterized by a different value of
$\Lambda^{2n}$ as in Eq.(\ref{scalesN}).
This will pave the way for the evaluation
of the coupling constants of the low-energy
$U(1)^n$ theory in section \ref{couplings}.
As is shown in \cite{Ferrari2}, the extremization of
the superpotential 
\be
{\cal W} = - \sum_{k=1}^C N_k \partial_{S_k} {\cal F}\,,  
\l{W}\ee
where the prepotential ${\cal F}$ is given by the planar
approximation to a holomorphic integral over
complex $n \times n$ matrices \cite{DV1,DV2,DV3}
\be
\exp \left( n^2 {\cal F}/S^2 \right)
= \int_{\rm planar} d^{n^2} \phi 
\exp \left[ - \frac{n}{S}\, tr {\cal W}_{tree}(\phi) \right]\,,
\l{F}\ee
is equivalent to solving the following equations
\be
\sum_{l=1}^C N_l t_{l k} = \,\,p_k\mod \,N_k\,,\, k=1,\ldots,C\,,  
\l{MMeom}\ee
where 
\be
t_{l k} = 
- \frac{1}{2 \pi i}\, \partial_{S_l} \partial_{S_k} {\cal F}\,,
\l{Ft}\ee
and $C$ is the number of cuts. 
In particular, $t_{l k}$ is the generalized period matrix  
$$
t_{l k} = 
\lim_{\ell_0 \to \infty} 
\left(
\frac{1}{2 \pi i} \int_{B_l} \psi_k\,dx\, 
+ \frac{i}{\pi} \log \ell_0 \,\right)\,,
$$
associated to the curve
$$
z^2 = \prod_{k=1}^C (x-a_k)(x-b_k)\,,
$$
which is the non-trivial part of the spectral curve
$$
y^2 = {\cal W}'_{tree}(x)^2 + f_{n-1}(x) 
= M(x)^2 \prod_{k=1}^C (x-a_k)(x-b_k)\,.
$$
The differentials $\psi_k\,dx\,$ form a basis
of log-normalizable holomorphic one-forms
dual to the contours $A_l$ surrounding the $C$ cuts,
namely they satisfy the following 
normalization condition \cite{Ferrari2} 
\be
\frac{1}{2 \pi i} \oint_{A_l} \psi_k\,dx = \delta_{k l}\,.
\l{normalpsi}\ee
It is possible to show that   
$$
\psi_k(x) = \frac{N_k(x)}{z}\,, 
$$
where 
$$
N_k(x) = x^{C-1} + ... \,,
$$
is a polynomial of degree $C-1$ whose coefficients are
determined by the normalization condition (\ref{normalpsi}).
Furthermore, the coupling constants of the 
$U(1)^C$ low-energy theory are given by
\be
\tau_{l k} = - \frac{1}{2 \pi i} 
\left(
\frac{\partial^2 {\cal F}}{\partial S_l \partial S_k}
- \delta_{l k} \frac{1}{N_l} \sum_{m=1}^C 
N_m \frac{\partial^2 {\cal F}}{\partial S_l \partial S_m}
\right)
= t_{l k} 
- \delta_{l k} \frac{1}{N_l} \sum_{m=1}^C t_{l m}\,,
\l{couplingC}\ee
and satisfy
$$
\sum_{k=1}^C \tau_{l k} N_k = 0\,,
$$
which signals the decoupling of the diagonal $U(1)$.

\subsection{The canonical basis}

\noindent
In our case $M(x) \equiv 1$ and 
we need to find a basis of one-forms
such that 
$$
\psi_k(x) = \frac{N_k(x)}{y}\,, 
$$
where 
$$
N_k(x) = x^{n-1} + ... \,,
$$
is a polynomial of degree $n-1$ whose coefficients are
determined by the normalization condition (\ref{normalpsi}).
In particular, the $\psi_k$'s will be a linear combination
of the one-forms $\omega_m = \frac{x^{m-1}}{y}\,dx\,, 
m=1,\ldots,n\,$
$$
\psi_k\,dx = \sum_{m=1}^n \alpha_{k m} \omega_m\,.
$$
In order to find the appropriate combination, 
we need to compute integrals of the form
\be
{\cal I}_{m l} \equiv \frac{1}{2 \pi i} \oint_{A_l} \omega_m \equiv 
\frac{1}{2 \pi i} \oint_{A_l} \frac{x^{m-1}}{y}\,dx\,, \quad l,m=1,\ldots,n\,. 
\l{calI}\ee
Then
$$
\frac{1}{2 \pi i} \oint_{A_l} \psi_k\,dx = 
\frac{1}{2 \pi i} \oint_{A_l} \alpha_{k m} \omega_m =
\alpha_{k m} {\cal I}_{m l} = \delta_{k l}\,, 
$$
which means that the matrix $\alpha$ is the inverse
of ${\cal I}$. It is convenient to consider 
integrals of the $\omega_m$'s
since, by virtue of $\ZZ_n$ the symmetry (\ref{Zn})
\be
{\cal I}_{m l} = e^{2 \pi i \frac{m}{n}} \,{\cal I}_{m\,l-1}
= \eta^{m} \,{\cal I}_{m\,l-1} \,.
\l{identity1}\ee
This identity will drastically simplify the analysis.
Define
$$
\hat{\cal I}_{k j} = \eta^{-(k-1)j} \,\hat{\cal I}_{1 j}\,,
\quad
\hat{\cal I}_{1 j} = \frac{1}{{\cal I}_{j 1}}\,. 
$$
Then
$$
\hat{\cal I}_{k j} \,{\cal I}_{j l} = 
\sum_{j=1}^n \eta^{-(k-1)j} \,\hat{\cal I}_{1 j}
\,\,\eta^{(l-1)j} \,{\cal I}_{j 1}
= \sum_{j=1}^n \eta^{(l-k)j} = n \delta_{l k}\,,
$$
which implies that 
$$
\alpha = {\cal I}^{-1} = \frac{1}{n} \hat{\cal I}\,. 
$$
The value of ${\cal I}_{n 1} = {\cal I}_{n j}$ can be evaluated 
using (\ref{identity1}) and
a simple residue calculation and it is given by
$$
n {\cal I}_{n 1} = \frac{1}{2 \pi i} \oint_{\sum A_l} \frac{x^{n-1}}{y} \,dx
= \frac{1}{2 \pi i} \oint_{A_{\infty}} \frac{x^{n-1}}{y} \,dx
= 1 \Rightarrow  {\cal I}_{n 1} = \frac{1}{n}\,. 
$$
Thus, we can verify that the basis one-forms
$\psi_k$ have the correct asymptotic behaviour
$$
\psi_k\,dx \sim \alpha_{k n} \frac{x^{n-1}}{y}\,dx 
= \frac{1}{n} \hat{\cal I}_{k n} \frac{x^{n-1}}{y}\,dx 
= \frac{1}{n} \hat{\cal I}_{1 n} \frac{x^{n-1}}{y}\,dx 
=  \frac{1}{n {\cal I}_{n 1}} \frac{x^{n-1}}{y}\,dx 
= \frac{x^{n-1}}{y}\,dx\,.
$$
In summary
\be
\psi_k\,dx = \frac{1}{n} 
\sum_{m=1}^n \eta^{-(k-1)m} \,\frac{\omega_m}{ {\cal I}_{m 1} }
\,.
\l{psik}\ee


\subsection{The period matrix}

The final goal is to evaluate
the generalized period matrix
$$
t_{l k} = t_{k l} \equiv \frac{1}{2 \pi i} \int_{B_l} \psi_k\,dx 
= \frac{1}{2 \pi i} \int_{B_l} \alpha_{k m} \omega_m 
= \frac{1}{n} \sum_{m=1}^n \hat{\cal I}_{k m} {\cal L}_{m l}\,, 
$$ 
where
\be
{\cal L}_{m l} \equiv \frac{1}{2 \pi i} \int_{B_l} \omega_m \,.
\l{calL}\ee
As before we find
\be
{\cal L}_{m l} = e^{2 \pi i \frac{m}{n}} \,{\cal L}_{m\,l-1}
= \eta^m {\cal L}_{m\,l-1}\,. 
\l{identity2}\ee

\noindent 
Then
$$
t_{l k} 
= \frac{1}{n} \sum_{m=1}^n \hat{\cal I}_{k m} {\cal L}_{m l}
= \frac{1}{n} \left( \sum_{m=1}^{n} \eta^{-m(k-1)} \hat{\cal I}_{1 m} 
\,\eta^{m(l-1)} {\cal L}_{m 1} \right)
$$
\be
= \frac{1}{n} \left( \sum_{m=1}^{n} \eta^{m(l-k)} \hat{\cal I}_{1 m} 
{\cal L}_{m 1} \right)
\equiv 
\frac{1}{n} \left( \sum_{m=1}^{n-1} \eta^{m(l-k)} c_{m} + c_n \right)\,.
\l{TlkFinal}\ee
Note that the above matrix is symmetric if and only if
$c_m = c_{n-m},\,m=1,\ldots,n-1$.
In fact  
$$
t_{l k} 
= \frac{1}{n} \left( \sum_{m=1}^{n-1} \eta^{m(l-k)} c_m 
+ c_n \right)
= t_{k l} 
= \frac{1}{n} \left( \sum_{m=1}^{n-1} \eta^{-m(l-k)} c_{m} 
+ c_n \right)
$$
$$
= \frac{1}{n} \left( \sum_{m=1}^{n-1} \eta^{(n-m)(l-k)} 
c_{m} + c_n \right)
= \frac{1}{n} \left( \sum_{m=1}^{n-1} \eta^{m (l-k)} 
c_{n-m} + c_n \right) 
$$
$$
\iff c_m = c_{n-m}\,, \quad m=1,\ldots,n-1\,. 
$$
This identity will be verified in section (\ref{checkMtoN-M}).

\subsection{The equations of motion}

The equations of motion (\ref{MMeom}) for $N_l = N$
reduce to
$$
\sum_{l=1}^n N_l t_{l k} = 
\sum_{l=1}^n N \,
\frac{1}{n} \left( \sum_{m=1}^{n} \eta^{m(l-k)} \hat{\cal I}_{1 m} 
{\cal L}_{m 1} \right)
$$
$$
= N \hat{\cal I}_{1 n} {\cal L}_{n 1}
= N n\, {\cal L}_{n 1}
= \,\,p_k\mod\,N
= \,\,p\mod\,N\,.
$$
Since $\omega_n$ is actually a logarithmic 
derivative
\be
\omega_n = \frac{x^{n-1}}{y}\,dx = 
\frac{1}{n} \frac{d}{dx}
\log \left( x^n - u + y
\right)\,,
\l{omegan}\ee
it follows that  
$$
{\cal L}_{n l} = \frac{1}{2 \pi i} \int_{B_l} \omega_n 
= \lim_{\ell_0 \to \infty} \frac{1}{\pi i} \int_{b_l}^{\ell_0} \omega_n 
- \frac{1}{\pi i} \log \ell_0
= - \frac{1}{n \pi i} \log (- \Lambda^{n})\,.
$$
where $b_l = e^{2 \pi i l/n} (u - 2 \Lambda^{n})^{1/n}$. 

\noindent
Therefore the equations of motion are equivalent to
$$
\sum_{l=1}^n N_l t_{l k} 
= - \frac{N}{\pi i} \log (- \Lambda^{n}) 
= \,\,p\mod \,N = \,\,- p'\mod \,N\,,
$$
which yields
\be
\Lambda^{2n} = e^{2 \pi i p'/N}\,,
\quad p'=0,1,\ldots,N-1\,. 
\l{Eom1}\ee
This reproduces the results of the strong coupling analysis 
(\ref{scalesN}).

\section{The $U(1)^n$ coupling constants}\label{couplings}

The coupling constant matrix $\tau_{l k}$ 
of the low-energy $U(1)^n$ theory is given by
\cite{DV1,DV2,DV3,CDSW}
\be
\tau_{l k} = 
t_{l k} 
- \delta_{l k} \frac{1}{N_l} \sum_{m=1}^n t_{l m}\,,
\l{coupling1}\ee
and satisfies $\sum_{k=1}^n \tau_{l k} N_k = 0$,
which signals the decoupling of the diagonal $U(1)$.

In Appendix B, we show that 
the periods of $\omega_m$ satisfy the 
following Picard-Fuchs equation
\be
\left(
\frac{\partial^2}{\partial u^2} 
+ \frac{\alpha(m)\,u }{(u^2-4\Lambda^{2n})} \frac{\partial}{\partial u}  
+ \frac{\beta(m)}{(u^2- 4\Lambda^{2n})} 
\right) \oint \omega_m = 0\,, 
\l{PFm2try}\ee\
where
\be
\alpha(m) = \frac{3n-2m}{n}\,, \quad \beta(m) = 
\left( \frac{\alpha(m)-1}{2} \right)^2 =
\left( \frac{n-m}{n} \right)^2\,.
\l{alphabeta}\ee
In terms of the variable $z = \frac{4 \Lambda^{2n}}{u^2}$,
Eq.(\ref{PFm2try}) is equivalent to
\be
\left( \frac{\partial^2}{\partial z^2}
+ \frac{3(1-z)-\alpha(m)}{2z(1-z)} \frac{\partial}{\partial z}  
+ \frac{\beta(m)}{4z^2(1-z)} \right) \phi(z) = 0\,. 
\l{PFm2}\ee
The {\it indicial equation} at $z=0$ has a double root 
equal to $(\alpha(m)-1)/4$.
This matches the behaviour of the integrals of $\omega_m$ 
around the cuts in the classical limit $z \to 0$
$$
\int_{A_k} \frac{x^{m-1}}{y} \,dx
\sim \oint_{x_k} \frac{x^{m-1}}{P_n(x)} 
\sim u^{\frac{m-n}{n}} 
= u^{(1-\alpha(m))/2} \sim 
z^{(\alpha(m)-1)/4}\,,
$$
where $x_k$ is the $k$-th root of $P_n(x)$. 

Setting $\phi(z) = z^{(\alpha-1)/4} \psi(z)$, we find
that (\ref{PFm2}) is equivalent to
the following hypergeometric equation for $\psi(z)$
\be
\left( 
z(1-z) \frac{\partial^2}{\partial z^2}  
+  \left( 1 - \frac{\alpha(m)+2}{2} z \right) \frac{\partial}{\partial z}  
- \frac{ \alpha^2(m)-1 }{16} \right) \psi(z) = 0\,, 
\l{PFfinalSC}\ee
with coefficients 
$$
a = \frac{\alpha(m)-1}{4} = \frac{n-m}{2n}\,,
\quad
b = \frac{\alpha(m)+1}{4} = \frac{2n-m}{2n}\,,
\quad
c = 1\,.
$$

\noindent
In terms of $w = 1 - z = 1 - \frac{4 \Lambda^{2n}}{u^2}$,
which is the appropriate variable in the neighbourhood
of the Argyres-Douglas point, Eq.(\ref{PFfinalSC})
becomes
\be
\left( w(1-w) \frac{\partial^2}{\partial w^2}  
+ \left( \frac{\alpha(m)}{2} - \frac{\alpha(m)+2}{2} w 
\right) 
\frac{\partial}{\partial w} 
- \frac{\alpha^2(m)-1}{16} 
\right) \psi(1-w) = 0\,, 
\l{PFfinalAD}\ee
which is a hypergeometric equation with
coefficients 
$$
a'= a\,, 
\quad
b'=b\,,
\quad 
c'=\frac{\alpha(m)}{2} = \frac{3n-2m}{2n}\,.
$$

\noindent
Thus, Eq.(\ref{PFfinalSC}) has two linearly independent
solutions, namely
\be
\psi_1(z) = F \left( \frac{n-m}{2n}, 
\frac{2n-m}{2n},1,z \right)\,,
\l{solution1}\ee
\be
\psi_2(z) = 
F \left( \frac{n-m}{2n}, 
\frac{2n-m}{2n}, \frac{3n-2m}{2n}, 1-z \right)\,.
\l{solution2}\ee


\noindent
We need to find the appropriate linear combination 
of $\psi_1(z)$ and $\psi_2(z)$ 
corresponding to each of the periods of $\omega_m$.
Let us first consider the integral 
of $\omega_m$ around the $k$-th branch cut
$$
\oint_{A_k} \omega_m = 
\oint_{A_k} \frac{x^{m-1}}{y}\,, m=1,\ldots,n\,. 
$$
By evaluating the above integral in the semiclassical 
limit $\frac{\Lambda^2}{u^2} \to 0$, we find
\be
\frac{1}{2 \pi i} \oint_{A_k} \frac{x^{m-1}}{y}
= \frac{1}{n} \eta^{(k-1)m} u^{\frac{m-n}{n}}\, 
\psi_1 \left( \frac{ 4 \Lambda^{2n} }{u^2} \right)\,,
\quad m=1,\ldots,n\,.
\l{Periods1}\ee

Eq. (\ref{TlkFinal}) reduces the calculation the  
period matrix $t_{l\,k}$ to the evaluation of the integrals
${\cal I}_{1 m}$ and ${\cal L}_{m 1}$
$$
{\cal I}_{1 m} \equiv \frac{1}{2 \pi i} \oint_{A_1} \omega_m\,,
\quad 
{\cal L}_{m 1} \equiv \frac{1}{2 \pi i} \int_{B_1} \omega_m\,.
$$
By virtue of (\ref{identity2}), the integral of $\omega_m$ along
the non-closed cycle $B_1$ can be related to
a period integral along a closed one. In fact  
$$
{\cal L}_{m 2} - {\cal L}_{m 1} =
\left( \eta^{m} - 1 \right) {\cal L}_{m 1} 
= \frac{1}{2 \pi i} \oint_{B_2 - B_1} \omega_m \,.
$$
Thus ${\cal L}_{1 m}$ will also be a solution of the
Picard-Fuchs equations derived above.
In the semiclassical limit, we find
$$
\frac{1}{2 \pi i} \oint_{B_2 - B_1} \frac{x^{m-1}}{y}\,dx 
= - \frac{2}{2 \pi i}
\int_{(u - 2\Lambda^n)^{1/n}}^{\,\eta (u + 2\Lambda^n)^{1/n}} 
\frac{x^{m-1}}{y}\,dx 
$$
$$
= - \frac{2}{2 \pi i}
u^{\frac{m-n}{n}} \frac{1}{n}
\left ( (\eta^m-1) \left( \gamma + \psi(m/n)  
+ \frac{1}{2} \log \left( \frac{ 4 \Lambda^{2n} }{u^2} \right) 
\right) + i \pi \eta^m \right)\, 
+ {\cal O}\left( \frac{4 \Lambda^{2n}}{u^2} \right)\,.
$$
This fixes the integral to be
\be
{\cal L}_{m1}   
= - \frac{2}{2 \pi i} u^{\frac{m-n}{n}} \frac{1}{n}
\left[ 
\,\log 2\,\, \psi_1 
- \frac{n}{2m-n} \frac{1}{A_2(m)} \left(\, 
\frac{\psi_1}{A_1(m)} -\psi_2 \right)\,
+ i \pi \frac{\eta^m}{\eta^m-1} \psi_1 \right]\,,
\l{NUnew}\ee
where
\be
A_1(m) = \frac{ \Gamma\left( (2m-n)/2n \right) }
{ \Gamma\left(m/2n \right) \Gamma\left( (n+m)/2n \right)}\,, \quad
A_2(m) = \frac{ \Gamma\left( (n-2m)/2n \right) }
{ \Gamma\left( (n-m)/2n \right) \Gamma\left( (2n-m)/2n \right)}\,.
\l{A1andA2}\ee
Thus
$$
c_m \equiv \frac{ {\cal L}_{m,1} }{ {\cal I}_{1m} } =
\frac{ \oint_{b_2 - b_1} \frac{x^{m-1}}{y}\,dx }{\eta^m-1} \left/
\oint_{A} \frac{ x^{m-1} }{ y } \,dx \right.
$$
\be
= - \frac{2}{2 \pi i} 
\left[ 
\,\log 2 - \frac{n}{2m-n} \frac{1}{A_1(m) A_2(m)} 
+ \frac{n}{(2m-n)A_2(m)} \frac{\psi_2}{\psi_1} \,
+ \frac{i \pi \eta^m}{(\eta^m-1)}
\right]\,.
\l{cmNEW}\ee

\subsection{The map $ m \to n - m $}\label{checkMtoN-M}

As was remarked above, the period matrix
(\ref{TlkFinal}) is symmetric if and only if
$c_{n-m}=c_m\,, m=1,\ldots,n-1$.
In Appendix C, 
it is shown that 
\be
\psi_1(n-m) 
= \left( 1 - \frac{ 4 \Lambda^{2n} }{u^2} \right)^{\frac{n-2m}{2n}} \psi_1(m)\,,
\l{tildePsi1}\ee
\be
\psi_2(n-m) 
= ( 1 - \frac{ 4 \Lambda^{2n} }{u^2} )^{\frac{n-2m}{2n}} 
\left( \frac{\psi_1(m)}{A_2(m)} - \frac{A_1(m)}{A_2(m)} \psi_2(m) 
\right)\,.
\l{tildePsi2}\ee
\noindent
Then
$$
c_{n-m} 
= - \frac{2}{2 \pi i} 
\left( \log 2
- \frac{n}{(n-2m)A_2(m)}
\frac{\psi_2(m)}{\psi_1(m)} 
+ i \pi \frac{ 1 }{ 1 - \eta^{m} }
\,\right)\,,
$$
and 
$$
c_{n-m} - c_m = 
- \frac{2}{2 \pi i} 
\left( \frac{n}{2m-n} \frac{1}{A_1(m) A_2(m)} 
+ i \pi \frac{ 1 }{ 1 - \eta^{m} }
- i \pi \frac{ \eta^m }{ \eta^{m} - 1 }
\,\right)\,
$$
\be
= - \frac{2}{2 \pi i} 
\left( \pi \cot\left( \frac{\pi m}{n} \right) 
+ i \pi \frac{ 1 + \eta^m }{ 1 - \eta^{m} }
\,\right)\,
= 0\,,
\l{cn-mcm}\ee
where we used the fact that
$$
A_1(m) A_2(m) = \frac{ n \tan(\pi m/n) }{ \pi (2m-n) }\,.
$$

\subsection{The diagonalized coupling constant matrix}

By Eq.(\ref{TlkFinal}) we can immediately see that 
the following 
$$
\left( v_{\pm p} \right)_k = \eta^{\pm p k}\,, \quad k =1,...,n\,,
p = 1,...,n\,, 
$$
are eigenvectors of $t$ with eigenvalues $c_p=c_{n-p}$\,.
In fact 
$$
\sum_{k=1}^n t_{l k} \left( v_{\pm p} \right)_k 
= \frac{1}{n} \sum_{k=1}^n \sum_{m=1}^{n} \eta^{m(l-k)} c_m \eta^{\pm p k}
= \frac{1}{n} \sum_{m=1}^{n} \sum_{k=1}^n \eta^{(\pm p - m)k} 
\eta^{m l} c_m
$$
\be
= c_p \eta^{\pm p l} = c_p \left( v_{\pm p} \right)_l\,,  
\l{eigen}\ee
where we used the fact that 
$$
\sum_{k=1}^n \eta^{(p - m)k} = n \delta_{p,m}\,,
\quad
\sum_{k=1}^n \eta^{(- p - m)k} = n \delta_{n-p,m}\,,
$$
and $c_{n-p}=c_p$. 
Taking linear combinations of $v_{\pm p}$ we obtain
two linearly independent real eigenvectors.
Thus we can immediately conclude that the period
matrix $t$ has $c_m\,, m=1,\ldots,n$ as its eigenvalues.
Due to the symmetry $c_{n-m}=c_m$, $m=1,\ldots,n-1$ 
they come in pairs.    
Finally, the matrix $\tau_{l k}$ 
of $U(1)$ couplings (\ref{coupling1}) 
becomes
\be
\tau_{l k} = t_{l k} - \delta_{l k} \left( \frac{1}{N_k}
\sum_{m=1}^n N_m t_{m k} \right)
= t_{l k} - \delta_{l k} c_n\,.
\l{tau}\ee
Therefore, the eigenvalues of $\tau$ are given by
$0$ and $\tau_m = c_m-c_n\,, m=1,\ldots,n-1$.

\subsection{Non-analytic behaviour close to the Argyres-Douglas points}

The hypergeometric function $\psi_1$ (\ref{solution1}) 
has the following analytical continuation 
$$
\psi_1 
= A_1(m) F \left( \frac{n-m}{2n}, \frac{2n-m}{2n}, \frac{3n-2m}{2n}, 
1-\frac{ 4 \Lambda^{2n} }{u^2} \right)
$$
\be
+ A_2(m) \left( 1 - \frac{ 4 \Lambda^{2n} }{u^2} \right)^{\frac{2m-n}{2n}}
F \left( \frac{n+m}{2n}, \frac{m}{2n}, \frac{n+2m}{2n}, 1 -\frac{ 4 \Lambda^{2n} }{u^2}
\right)\,, \quad m=1,\ldots,n-1\,, 
\l{AC1}\ee
except for the case $m = \frac{n}{2}$ when
\be
\psi_1 = \frac{1}{\sqrt{2} \pi}
\sum_{p=0}^\infty \frac{(\frac{1}{4})_p (\frac{3}{4})_p}{p! p!}
\left[ k_p - \log\left(  1 - \frac{ 4 \Lambda^{2n} }{u^2}  
\right)\, \right] \left(  1 - \frac{ 4 \Lambda^{2n} }{u^2}  \right)^p\,,
\l{AC2}\ee
where $k_p = 2\,\psi(p+1)-\psi\left(\frac{1}{4} + p\right)
-\psi\left(\frac{3}{4}+p\right)$.
\noindent
Then, by (\ref{cmNEW}), the eigenvalues of the period
matrix are non-analytic at the Argyres-Douglas points.
In fact
\be
c_m = c_{n-m} \approx 
\left(  1 - \frac{ 4 \Lambda^{2n} }{u^2}  \right)^{\frac{n-2m}{2n}}\,,
\quad m < \frac{n}{2}\,, 
\quad
c_{\frac{n}{2}} \approx  
\frac{1}{\log \left( 1 - \frac{ 4 \Lambda^{2n} }{u^2} \right)}\,.
\l{asymptAD}\ee

\subsection{Transition to a solution with
a lower number of cuts}

By (\ref{Periods1}),(\ref{AC1}) and (\ref{AC2}) 
$$
\lim_{u^2 \to 4 \Lambda^{2n}} {\cal I}_{m,l} 
\equiv \frac{1}{2 \pi i} \oint_{\gamma_l} \frac{x^{m-1}}{y}\,dx
\to \infty  \,, \quad m \leq \frac{n}{2}\,.
$$
Then 
\be
\lim_{u^2 \to 4 \Lambda^{2n}} \psi_k 
= \sum_{m > n/2} \frac{1}{{\cal I}_{mk}} \omega_m\,.
\l{limitpsi}\ee
Note that in this limit
$$
\lim_{u \to \pm 2\Lambda^{n}} \omega_m 
= \frac{x^{m-1}}{ \sqrt{ x^n(x^n \mp 4 \Lambda^{2n}) } }\,dx\,.
$$
Therefore, for $n=2p$ and $m > \frac{n}{2} = p\,$ we find
\be
\lim_{u \to \pm 2\Lambda^{n}} \omega_m 
= \frac{x^{m-1-p}}{ \sqrt{(x^{2p} \mp 4 \Lambda^{2n}) } }\,dx
= \frac{x^{m'-1}}{ \tilde y }\,dx \equiv
\tilde \omega_{m'}\,, \quad m'=1,...,p\,.
\l{neven}\ee
Likewise, for $n=2p+1$, $m > p$ 
\be
\lim_{u \to \pm 2\Lambda^{n}} \omega_m 
= \frac{x^{m-1-p}}{ \sqrt{x (x^{2p+1} \mp 4 \Lambda^{2n}) } }\,dx
= \frac{x^{m'-1}}{ \tilde y }\,dx 
\equiv \tilde \omega_{m'}\,
\,,
\quad
m'=1,...,p+1\,.
\l{nodd}\ee
Hence, by (\ref{limitpsi}),(\ref{neven}) and (\ref{nodd}),
we can conclude that in the Argyres-Douglas limit
the $n$-cut solution degenerates into 
one with $\frac{n}{2}$ cuts for $n$ even and 
$\frac{n+1}{2}$ cuts for $n$ odd, in short
an $\left[ \frac{n+1}{2} \right]$-cut solution. 
The relevant curves are 
given respectively by
$$
\tilde y^2 = x^{n} \mp 4 \Lambda^{2n}\,,
$$
and
$$
\tilde y^2 =  x (x^{n} \mp 4 \Lambda^{2n}) \,.
$$
The above generalizes a result of \cite{Ferrari2}, where 
the singularity corresponded to a transition
from a two-cut solution to a one-cut solution.


\section{The large $N$ limit}\l{largeN}

Following \cite{Ferrari2,FerrariNCST}, we expect 
a non-trivial behaviour of the large $N$ limit 
at the Argyres-Douglas critical points.
Using Eqs.(\ref{cmNEW}) and (\ref{tau}),
we can analyze the behaviour of the coupling
constants of the low-energy $U(1)^n$ theory
in the $p$-th vacuum, $p=1,\ldots,N$.  
Let us denote by $\tau_m$, $m=1,\ldots,n-1$  
the non-trivial eigenvalues of the couling constant
matrix $\tau_{lk}$. 
We find that
$$
\tau_m = c_m(\,e^{2\pi i p/N}\,x\,) - c_n =
\frac{p}{N} + c_m(\,e^{2\pi i p/N}\,x\,) 
$$
\be
=
\frac{p}{N} + c_m(x) + \frac{2 \pi i p}{N} x\, c_m'(x)  
+ {\cal O}\left(\frac{1}{N^2}\right)\,,
\l{tauMfirstexp}\ee
where we used the fact that by Eq.(\ref{Eom1})
$$
\frac{4\Lambda^{2n}}{u^2} = 
\frac{ 4 e^{2\pi i p/N} }{u^2}
\equiv e^{2\pi i p/N} x\,, 
\quad p=1,\ldots,N\,.
$$
\noindent
However, the expansion (\ref{tauMfirstexp}) 
is singular in the vicinity of the Argyres-Douglas points,  
$x=x_c=1$, because by (\ref{asymptAD}), 
$c_m'(x)$ is not defined for $x=x_c$.
Actually, at the Argyres-Douglas points, 
$\tau_m$ becomes
\be
\tau_m \approx \left( \frac{p}{N} \right)^{\frac{n-2m}{2n}}\,, 
\quad m < \frac{n}{2}\,, 
\quad
\tau_{\frac{n}{2}} \approx   
\frac{1}{\log \left( \frac{p}{N} \right)}\,.
\l{tauMexactx=1}\ee
This is clearly  
a signal of the breakdown of the large $N$ expansion.

\noindent
In fact, the $1/N$ corrections to
(\ref{tauMfirstexp}) for $x \ne x_c$ read
\be
\tau_m \approx (1-x)^{\frac{n-2m}{2n}}
\left[
- 2\pi i \,\left( \frac{n-2m}{2n} \right) \frac{p}{N(1-x)} 
+ {\cal O}\left( \frac{1}{(N(1-x))^2} \right)
\right]\,, m < \frac{n}{2}\,, 
\l{tauMN(1-x)}\ee
\be
\frac{1}{\tau_{ \frac{n}{2}} } \approx 
\log(1-x) - 2\pi i \frac{p}{N(1-x)} 
+ {\cal O}\left( \frac{1}{(N(1-x))^2} \right) 
\l{tauMN(1-x)2}\ee
which make the singularity manifest.

\subsection{The double scaling limit}

Eqs. (\ref{tauMN(1-x)}) and (\ref{tauMN(1-x)2})
suggest that the divergences at 
$x=1$ can be compensated by taking the limits 
$N \to \infty$ and $x \to x_c=1$ in a correlated way
as follows
\be
x \to 1 \,, \quad N \to \infty\,, 
\quad  N(1-x) = cnst = \frac{1}{\kappa}\,.
\label{double}\ee

\noindent 
In particular, the rescaled couplings 
\be
\tau^{scaled}_m = (1-x)^{\frac{2m-n}{2n}} \tau_m\,,
\l{tauscaled}\ee
have a finite universal limit given by 
\be
\tau^{scaled}_m \sim
(1 - 2 \pi i p \kappa)^{\frac{n-2m}{2n}}\,.
\label{limittauscaled}\ee 
Similarly, for $1/\tau_{n/2}$, after
subtracting a term proportional to
$\log(1-x)$, one obtains
\be                                                                      
\frac{1}{\tau^{scaled}_{n/2}} \sim
\log \left( 1 - 2 \pi i p \kappa \right)\,.
\label{limittauscaled2}\ee 

\noindent 
In a series of papers \cite{FerrariNCST,FerrariNPB612,
FerrariNPB617,FerrariNCSTlast}, Ferrari 
made a proposal to generalize the matrix model
approach to non-critical strings \cite{2dMMNCST}
to the four dimensional case. 
The basic idea
is to replace matrix integrals with four dimensional
gauge theory path integrals with $N \times N$
adjoint Higgs fields. 

It was shown in \cite{FerrariNPB612} 
that the large $N$ expansion of pure 
${\cal N}=2$ supersymmetric $SU(N)$ gauge theory
becomes singular at special points on the moduli
space due to IR divergences. 
However, these divergences
can be compensated by taking the limit $N \to \infty$
and approaching the critical points in
a correlated manner. These {\it double scaling
limits} were then conjectured to define 
four dimensional string theories \cite{FerrariNPB617}. 
In this paper, the Seiberg-Witten period
integrals of the $A_{n-1}$ Argyres-Douglas singularities 
were analyzed in detail.
It is crucial to note there are 
non-trivial contributions in powers 
of $1/N$, which signals the presence of opens strings. 
These terms are generated by 
fractional instantons \cite{FerrariNPB612}.

In \cite{Ferrari2}, the analysis was extended
to the ${\cal N}=1$ case. In particular, a 
$U(N)$ gauge theory with cubic superpotential
was studied and it was shown that there
are critical values of the superpotential
couplings where glueballs are massless,
there are tensionless domain walls and
confinement without a mass gap. At these
critical points, the large $N$ expansion is singular 
and the tension of domain walls scales 
as a fractional power of $N$. Nevertheless, 
double scaling limits analogous to  
(\ref{double}) exist and are again
conjectured to define a four dimensional 
non-critical string theory. 

The double scaling limits (\ref{double})
fit into the above scenario and are
consistent with the ${\cal N}=2$
analysis of \cite{FerrariNPB617}.
The conjecture is that they define a
four dimensional non-critical string theory.

\section{Conclusion}\label{Conclu}

Using the techniques of \cite{CIV},
we constructed an ${\cal N}=1$ theory with gauge group
$U(nN)$ and degree $n+1$ tree level superpotential
whose matrix model spectral curve develops
an $A_{n-1}$ Argyres-Douglas singularity. This
theory is closely related to an underlying ${\cal N}=2$
$U(n)$ model. In fact, the one-dimensional 
parameter space of the $U(nN)$ theory is actually
isomorphic to a slice of the ${\cal N}=2$ 
Coulomb moduli space of the $U(n)$ theory:
$n-1$ parameters of the 
$U(n)$ Seiberg-Witten curve are set to zero
and the remaining one parametrizes
the most relevant deformation away
from the singularity.
In particular, only
a finite, $N$-independent, number of parameters
is adjusted. 
The expression of the coupling constants
of the $U(1)^n$ low-energy theory
shows that the $1/N$
expansion is singular at the Argyres-Douglas points.
Nevertheless, it is possible to define
appropriate double scaling limits (\ref{double})
which are conjectured to define
four dimensional
non-critical string theories as proposed by
Ferrari in 
\cite{FerrariNCST,FerrariNPB612,FerrariNPB617,FerrariNCSTlast,FerrariLH}. 
At the Argyres-Douglas points, the $n$-cut matrix model spectral
curve degenerates into a curve with $\frac{n}{2}$ cuts for $n$ even
and $\frac{n+1}{2}$ cuts for $n$ odd.

\section*{Acknowledgements}

I would like to thank Freddy Cachazo, Frank Ferrari,
Amihay Hanany, Marco Matone
and David Tong  
for discussions and useful comments.
I would like to thank Freddy Cachazo for pointing
out a mistake in a previous version of the paper.
My work is supported in part by 
the CTP and the LNS of MIT, by the U.S. Department of Energy
under cooperative research agreement \# DE-FC02-94ER40818,
by the INFN ``Bruno Rossi'' Fellowship and the 
Foundation BLANCEFLOR Boncompagni-Ludovisi, n\'ee Bildt.



\section*{Appendix A}

Performing the analytic continuation of (\ref{Szero}) we find
$$
S_k = C_{1,k} \, F\left( -\frac{1}{2} -\frac{1}{2n}, 
\frac{1}{2} -\frac{1}{2n}, \frac{1}{2}, \frac{u^2}{4 \Lambda^{2n}} 
\right) 
+ C_{2,k} \, u\, F\left( - \frac{1}{2n}, -\frac{1}{2n} + 1, \frac{3}{2},\frac{u^2}{4 \Lambda^{2n}} \right)
$$
$$
= \frac{C_{1,k} \Gamma(1/2)}{\Gamma(1/2 - 1/2n)} 
\left[ \frac{(-u^2/4 \Lambda^{2n})^{-1/2+1/2n}}{\Gamma(1 + 1/2n)} \sum_{k=0}^{\infty}
\left[
\frac{(-1/2-1/2n)_{k+1} (-1/2n)_{k+1}}{k!(k+1)!} \left( \frac{4 \Lambda^{2n}}{u^2}\right)^{k}
\right. \right.
$$
$$
\left.
\left.
 \times \left( \log(-u^2/4\Lambda^{2n}) + h_{1,k} \right) \right]
+ (-u^2/4 \Lambda^{2n})^{+1/2+1/2n} \frac{\Gamma(1)}{\Gamma(1 + 1/2n)}
\right] 
$$
$$
+ \frac{C_{2,k} \Gamma(3/2)}{\Gamma(1 - 1/2n)}\,u 
\left[ \frac{(-u^2/4 \Lambda^{2n})^{-1+1/2n}}{\Gamma(3/2 + 1/2n)} \sum_{k=0}^{\infty}
\left[
\frac{(-1/2-1/2n)_{k+1} (-1/2n)_{k+1}}{k!(k+1)!} \left( \frac{4 \Lambda^{2n}}{u^2}\right)^{k}
\right. \right.
$$
$$
\left.
\left.
 \times \left( \log(-u^2/4 \Lambda^{2n}) + h_{2,k} \right) \right]
+ (-u^2/4\Lambda^{2n})^{1/2n} \frac{\Gamma(1)}{\Gamma(3/2 + 1/2n)}
\right]\,. 
$$
In the above expression there is a term proportional to $u^{1+1/n}$ that
we need to set to zero.
This determines $C_{2,k}$ as a function of $C_{1,k}$
\be
C_{2,k} = -\frac{e^{i\pi/2}}{2 \Lambda^n} \frac{\Gamma(1-1/2n)\Gamma(3/2+1/2n)}{\Gamma(1/2 - 1/2n)\Gamma(1+1/2n)}\,C_{1,k}\,.
\l{C2}\ee
\noindent
We are left with
$$
S_k = \frac{C_{1,k} \Gamma(1/2)}{\Gamma(1/2 - 1/2n)} 
\frac{(-u^2/4\Lambda^{2n})^{-1/2+1/2n}}{\Gamma(1 + 1/2n)} 
$$
$$
\sum_{k=0}^{\infty}
\frac{(-1/2-1/2n)_{k+1} (-1/2n)_{k+1}}{k!(k+1)!} 
\left( \frac{4\Lambda^{2n}}{u^2} \right)^{k}
\left( h_{1,k} - h_{2,k} \right) 
$$
where
$$
h_{1,k} - h_{2,k} 
= -\psi(1/2 -1/2n + k) - \psi(1/2n - k)
+ \psi(-1/2n + 1 + k) + \psi(1/2 +1/2n - k)
$$
$$
=
\left[ \psi(1/2 +1/2n - k) -\psi(1/2 -1/2n + k) \right] 
- \left[ \psi(1/2n - k) - \psi(-1/2n + 1 + k) \right]
$$
$$
=
\left[ \pi \tan \pi(1/2n - k) \right] 
- \left[ - \pi \cot \pi(1/2n - k) \right] = \frac{2 \pi}{\sin(\pi/n)}\,.
$$
Finally
$$
S_k = \frac{2 \pi}{\sin(\pi/n)}\,\frac{C_{1,k} \Gamma(1/2)}{\Gamma(1/2 - 1/2n)} 
\frac{(-u^2/4\Lambda^{2n})^{-1/2+1/2n}}{\Gamma(1 + 1/2n)} \times
$$
$$
\times
\sum_{k=0}^{\infty}
\frac{(-1/2-1/2n)_{k+1} (-1/2n)_{k+1}}{k!(k+1)!} 
\left( \frac{4\Lambda^{2n}}{u^2} \right)^{k}\,.
$$
Using
$$
(-1/2-1/2n)_{k+1} \equiv \frac{\Gamma(-1/2-1/2n + k + 1)}{\Gamma(-1/2-1/2n)}
=  \frac{\Gamma(1/2-1/2n + k)}{\Gamma(1/2-1/2n)}\, \frac{\Gamma(1/2-1/2n)}{\Gamma(-1/2-1/2n)}
$$
$$
= (1/2-1/2n)_{k}\, \frac{\Gamma(1/2-1/2n)}{\Gamma(-1/2-1/2n)}\,,
$$
$$
(-1/2n)_{k+1} \equiv \frac{\Gamma(-1/2n + k + 1)}{\Gamma(-1/2n)}
=  \frac{\Gamma(1-1/2n + k)}{\Gamma(1-1/2n)} \,
\frac{\Gamma(1-1/2n)}{\Gamma(-1/2n)}
$$
$$
= (1-1/2n)_{k}\, \frac{\Gamma(1-1/2n)}{\Gamma(-1/2n)}\,,
$$
and
$$
\Gamma(2) (2)_k = (2)_k = \Gamma(2+k) =  (k+1)! 
$$
the expression is equivalent to 
$$
S_k = \frac{2 \pi}{\sin(\pi/n)}\,\frac{C_{1,k} \Gamma(1/2)}{\Gamma(1/2 - 1/2n)} 
\frac{(-u^2/4\Lambda^{2n})^{-1/2+1/2n}}{\Gamma(1 + 1/2n)} 
\frac{\Gamma(1/2-1/2n)}{\Gamma(-1/2-1/2n)}
\frac{\Gamma(1-1/2n)}{\Gamma(-1/2n)}
$$
$$
\sum_{k=0}^{\infty}
\frac{(1/2-1/2n)_{k} (1-1/2n)_{k}}{k!\, (2)_{k}} 
\left( \frac{4\Lambda^{2n}}{u^2} \right)^{k} 
$$
$$
= \frac{2 \pi}{\sin(\pi/n)}\,\frac{ \sqrt{\pi} C_{1,k} }{\Gamma(1+1/2n)} 
\frac{(-u^2/4\Lambda^{2n})^{-1/2+1/2n}}{\Gamma(-1/2 - 1/2n)} 
\frac{\Gamma(1-1/2n)}{\Gamma(-1/2n)}  
$$
$$
F\left( 1/2 - 1/2n, 1 - 1/2n, 2, \frac{4\Lambda^{2n}}{u^2} \right)
$$
Comparing with (\ref{Sinfinity}), we find
\be
C_{3,k} = e^{i\pi/2 + i\pi/2n} \frac{2 \pi}{\sin(\pi/n)}\,
\left( 
\frac{\sqrt{\pi}\,2^{\frac{n-1}{n}} }{2n \Gamma(1 + 1/2n)\Gamma(-1/2-1/2n)}
\right) \Lambda^{n-1} C_{1,k}\,.
\label{C3}\ee

\section*{Appendix B: The Picard-Fuchs
equations for the coupling constants}
\label{PFcc}

In order to determine the
coupling constants explicitly, 
we will derive the Picard-Fuchs
equations satisfied by the periods of $\omega_m$.
The first derivatives of a period w.r.t. $u$
and $\Lambda^{2n}$ are given by
\be
\frac{\partial}{\partial u} \oint \omega_m =
\frac{\partial}{\partial u} \oint \frac{x^{m-1}}{y}\,dx 
= \oint \frac{ x^{n+m-1} - u x^{m-1} }{y^3}\,dx\,,  
\l{firstDu}\ee
and 
\be
\frac{\partial}{\partial \Lambda^{2n}} 
\oint \frac{x^{m-1}}{y}\,dx 
= 2 \oint \frac{x^{m-1}}{y^3}\,dx\,.  
\l{firstDL}\ee
Furthermore, since a period of $\omega_m$
is a homogeneous function of degree $m-n$,
we find
\be
\left( 2n \Lambda^{2n} 
\frac{\partial}{\partial \Lambda^{2n}}
+ n u \frac{\partial}{\partial u} - (m-n)
\right) \oint \frac{x^{m-1}}{y}\,dx 
= 0\,.
\l{homoge}\ee

\noindent
The second derivative w.r.t. $u$ reads 
$$
\frac{\partial^2}{\partial u^2} \oint \frac{x^{m-1}}{y}\,dx 
= 12 \Lambda^{2n} \oint \frac{x^{m-1}}{y^5}\,dx + 2 \oint \frac{x^{m-1}}{y^3}\,dx\,.  
$$
Then
$$
\frac{x^{m-1}}{y^5}\,dx 
=  
- \left( \frac{ 4 \Lambda^{2n} ( m -3n ) + m u^2}
{12 \,n\, \Lambda^{2n} (u^2 - 4 \Lambda^{2n})}
\right) \frac{x^{m-1}}{y^3}\,dx \,
-
\left( \frac{ ( m - 2n ) u }
{12 \,n\, \Lambda^{2n} (u^2 - 4 \Lambda^{2n})}
\right) \frac{x^{m+n-1}}{y^3}\,dx \,,
$$
up to a total derivative,
which yields
$$
\frac{\partial^2}{\partial u^2} \oint \frac{x^{m-1}}{y}\,dx 
=
\frac{(m-2n)u}{n(u^2 - 4 \Lambda^{2n})} \oint \frac{x^{n+m-1}}{y^3}\,dx 
$$
\be
- \frac{\left( 4 \Lambda^{2n} (m-n) + (m-2n) u^2 
\right)}{n(u^2- 4 \Lambda^{2n})}  
\oint \frac{x^{m-1}}{y^3}\,dx\,.
\l{prep2}\ee
Note that all the above equations
involve periods of $\frac{x^{m-1}}{y^3}\,dx$
and $\frac{x^{n+m-1}}{y^3}\,dx\,$ only.
Inverting these relations yields
the Picard-Fuchs equation for the periods of $\omega_m$
\be
\left(
\frac{\partial^2}{\partial u^2} 
+ \frac{\alpha(m)\,u }{(u^2-4\Lambda^{2n})} \frac{\partial}{\partial u}  
+ \frac{\beta(m)}{(u^2- 4\Lambda^{2n})} 
\right) \oint \omega_m = 0\,, 
\l{PFm2tryAPP}\ee
\be
\alpha(m) = \frac{3n-2m}{n}\,, \quad \beta(m) = 
\left( \frac{\alpha(m)-1}{2} \right)^2 =
\left( \frac{n-m}{n} \right)^2\,.
\l{alphabetaAPP}\ee

\section*{Appendix C}\label{mTOn-m}

First of all
$$
\psi_1(n-m) = F \left( a(n-m), b(n-m), c(n-m), 
\frac{ 4 \Lambda^{2n} }{u^2} 
\right)
$$
$$ 
= F \left( c(m) - b(m) , c(m) - a(m) , c(m), \frac{ 4 \Lambda^{2n} }{u^2} 
\right) 
$$
$$
= \left( 1 - \frac{ 4 \Lambda^{2n} }{u^2} \right)^{a(m)+b(m)-c(m)}
\psi_1(m) 
= \left( 1 - \frac{ 4 \Lambda^{2n} }{u^2} \right)^{\frac{n-2m}{2n}}
\psi_1(m)\,,
$$
where the following Kummer's relation was used
$$
F(a,b,c,z) = (1-z)^{c-a-b} F(c-a,c-b,c,z)\,.
$$
Likewise
$$ 
\psi_2(n-m) = F \left( a(n-m), b(n-m), c(n-m), 1-\frac{ 4 \Lambda^{2n} }{u^2}  
\right)
$$
$$ 
= F \left( 1 - a(m) , 1 - b(m) , 2 - c(m), 1- \frac{ 4 \Lambda^{2n} }{u^2} 
\right) 
$$
$$
= \left( 1 - \frac{ 4 \Lambda^{2n} }{u^2} \right)^{c(m)-1} \,
\left( \frac{ 4 \Lambda^{2n} }{u^2} \right)^{a(m)+b(m)-c(m)} 
{\cal U}_5\left( 1 - \frac{ 4 \Lambda^{2n} }{u^2} \right)
$$
$$
=
\left( 1 - \frac{ 4 \Lambda^{2n} }{u^2} \right)^{c(m)-1} \, 
{\cal U}_5\left( 1 - \frac{ 4 \Lambda^{2n} }{u^2} \right)
$$
Another Kummer's relation
$$
{\cal U}_5( 1-z )
= \frac{\Gamma(c-a)\Gamma(c-b)}{\Gamma(c+1-a-b)\Gamma(c-1)}
{\cal U}_6( 1-z )
-\frac{\Gamma(c-a)\Gamma(c-b)\Gamma(1-c)}{\Gamma(c-1)\Gamma(1-a)\Gamma(1-b)} {\cal U}_1( 1-z )
$$
$$
= \frac{\Gamma(c-a)\Gamma(c-b)}{\Gamma(c+1-a-b)\Gamma(c-1)}
( z )^{c-a-b} F(c-a,c-b,c+1-a-b,z)
$$
$$ 
-\frac{\Gamma(c-a)\Gamma(c-b)\Gamma(1-c)}{\Gamma(c-1)\Gamma(1-a)\Gamma(1-b)} F(a,b,c,1-z)\,,
$$
implies that
$$
\psi_2(n-m) = 
\left( 1 - \frac{ 4 \Lambda^{2n} }{u^2} \right)^{c(m)-1} \, 
{\cal U}_5\left( 1 - \frac{ 4 \Lambda^{2n} }{u^2} \right)
$$
$$
= \left( 1 - \frac{ 4 \Lambda^{2n} }{u^2} \right)^{c(m)-1} 
\left( \frac{\Gamma(c-a)\Gamma(c-b)}{\Gamma(c-1)} \psi_1(m) 
- \frac{\Gamma(c-a)\Gamma(c-b)\Gamma(1-c)}{\Gamma(c-1)\Gamma(1-a)\Gamma(1-b)} 
 \psi_2( m ) \right)
$$
$$
=
\left( 1 - \frac{ 4 \Lambda^{2n} }{u^2} \right)^{\frac{n-2m}{2n}} 
\left( \frac{\Gamma((n-m)/2n)\Gamma((2n-m)/2n)}{\Gamma((n-2m)/2n)} \psi_1(m) 
\right.
$$
$$
\left.
- \frac{\Gamma((n-m)/2n)\Gamma((2n-m)/2n)\Gamma((2m-n)/2n)}
{\Gamma(m/2n)\Gamma((n+m)/2n)\Gamma((n-2m)/2n)} 
 \psi_2( m ) \right)
$$
$$
= \left( 1 - \frac{ 4 \Lambda^{2n} }{u^2} \right)^{\frac{n-2m}{2n}} 
\left( \frac{\psi_1(m)}{A_2(m)} - \frac{A_1(m)}{A_2(m)} \psi_2(m) 
\right)\,.
$$
\noindent

\section*{Appendix D}

In order to derive the various Picard-Fuchs equations
in the paper, we make use of the following identity
\be
a(x) p(x) + b(x) p'(x) = u^2 - 4\Lambda^{2n}\,, 
\l{identityPF1}\ee
where
$$
p(x) = y^2 = (x^n-u)^2 - 4 \Lambda^{2n}\,,  
$$
and
$$
a(x) = 1 - \frac{u}{4 \Lambda^{2n}} x^n\,,
\quad
b(x) = \frac{x}{2n} \left( -1 + \frac{u}{4 \Lambda^{2n}} (x^n-u) 
\right)\,.
$$
By (\ref{identityPF1}), we see that 
$$
\frac{\phi(x)}{y^n} =
\frac{1}{u^2 - 4 \Lambda^{2n}} 
\left(
\frac{a(x) \phi(x)}{y^{n-2}} + 
\frac{2}{n-2} \,\frac{ ( \,b(x) \phi(x) )' }{y^{n-2}}\,
\right)\,,
$$
up to a total derivative.

\bibliographystyle{JHEP}

\end{document}